\documentclass[%
 reprint,
superscriptaddress,
nofootinbib,
 amsmath,amssymb,
 aps,
]{revtex4-1}

\usepackage{graphicx}% Include figure files
\usepackage{dcolumn}% Align table columns on decimal point
\usepackage{bm}% bold math

\usepackage{comment}

\begin{document}

\preprint{APS/123-QED}

\title{Detecting Intermediate-Mass Ratio Inspirals From The Ground And Space}%\\
       %With Ground-Based Gravitational Wave Detectors}

\author{Pau Amaro-Seoane}
\affiliation{Institute of Space Sciences (ICE, CSIC) \& Institut d'Estudis Espacials de Catalunya (IEEC) at Campus UAB, Carrer de Can Magrans s/n 08193 Barcelona, Spain}
\affiliation{Institute of Applied Mathematics, Academy of Mathematics and Systems Science, Chinese Academy of Sciences, Beijing 100190, China}
\affiliation{Kavli Institute for Astronomy and Astrophysics at Peking University, 100871 Beijing, China}
\affiliation{Zentrum f{\"u}r Astronomie und Astrophysik, TU Berlin, Hardenbergstra{\ss}e 36, 10623 Berlin, Germany
}

\date{\today}

\begin{abstract}
The detection of a gravitational capture of a stellar-mass compact object by a
massive black hole (MBH) will allow us to test gravity in the strong regime.
The repeated, accumulated bursts of gravitational radiation from these sources
can be envisaged as a geodesic mapping of space-time around the MBH.
These sources form via two-body relaxation, by exchanging energy and angular
momentum, and inspiral in a slow, progressive way down to the final merger. The
range of frequencies is localised in the range of millihertz in the case of MBH
of masses $\sim 10^6\,M_{\odot}$, i.e. that of space-borne gravitational-wave
observatories such as LISA.
In this article I show that, depending on their orbital parameters, intermediate-mass ratios (IMRIs) of MBH of masses between a hundred
and a few thousand have frequencies that make them detectable (i) with
ground-based observatories, or (ii) with both LISA and ground-based ones such
as advanced LIGO/Virgo and third generation ones, with ET as an example.  The
binaries have a signal-to-noise ratio large enough to ensure detection. More
extreme values in their orbital parameters correspond to systems detectable
only with ground-based detectors and enter the LIGO/Virgo band in particular in
many different harmonics for masses up to some $2000,\,M_{\odot}$. I show that
environmental effects are
negligible, so that the source should not have this kind of complication. The accumulated phase-shift is measurable with LISA and ET, and for some cases also with LIGO, so that it is
possible to recover information about the eccentricity and formation scenario.
For IMRIs with a total mass $\lessapprox 2000\,M_{\odot}$ and initial
eccentricities up to $0.999$, LISA can give a warning to ground-based detectors
with enough time in advance and seconds of precision. The possibility of
detecting IMRIs from the ground alone or combined with space-borne
observatories opens new possibilities for gravitational wave astronomy.

\end{abstract}

                             % Classification Scheme.
                              %display desired
\maketitle

\section{Introduction}\label{sec:intro}

The typical size of a massive black hole (MBH), i.e. its Schwarzschild radius,
is from the point of view of the host galaxy extremely tiny. For a
$10^6\,M_{\odot}$ MBH, this difference spans over ten orders of magnitude.
However, we have discovered a deep link between the properties of the galaxy
and those of the MBH, in particular between the mass of the MBH and the
velocity dispersion $\sigma$ of the spheroidal component of the galaxy
\citep{KormendyHo2013}. Because the region of interest is difficult to resolve,
the lower end of this correlation is uncertain. However, if we extend these
correlations to smaller systems, globular clusters, or ultra-compact dwarf
galaxies should harbour black holes with masses ranging between $10^2$ and
$10^4,\,M_{\odot}$, i.e. intermediate-mass black holes, IMBHs \citep[for a
review, see the work of][]{Mezcua2017,LuetzgendorfEtAl2013}, although
such black holes have never been robustly detected.

The best way to probe the nature of the MBH is with gravitational waves, which
allow us to extract information that is unavailable electromagnetically. The
gravitational capture and plunge of a compact object through the event horizon
is one of the main goals of the Laser Interferometer Space Antenna (LISA)
mission \citep{Amaro-SeoaneEtAl2017}. A compact object of stellar mass, so
dense that it defeats the tidal forces of the MBH, is able to approach very
closely the central MBH, emitting a large amount of gravitational radiation as
orbital energy is radiated away. This causes the semi-major axis to shrink.
This ``doomed'' object spends many orbits around the MBH before it is
swallowed. The radiated energy which can be thought of as a snapshot containing
detailed information about the system will allow us to probe strong-field
gravitational physics. Depending on the mass ratio $q$, we talk about either
extreme-mass ratio inspirals, $q \gtrsim 10^4:1$ (EMRI,
see \cite{Amaro-SeoaneLRR2012,Amaro-SeoaneGairPoundHughesSopuerta2015}) or
intermediate-mass ratio inspirals, $q \sim 10^2-10^4:1$ (IMRI, see
e.g. \cite{Amaro-SeoaneEtAl07,BrownEtAl2007,RodriguezEtAl2012}).

In galactic nuclei the predominant way of producing EMRIs is via two-body
relaxation \citep{Amaro-SeoaneLRR2012}. At formation, these sources have
extremely large eccentricities, particularly if the MBH is Kerr
\citep{Amaro-SeoaneSopuertaFreitag2013}, which is what we should expect from
nature. However, in globular clusters, which harbour MBH in the range of IMBHs,
the loss-cone theory, which is our tool to understand how EMRIs form,
\citep[see e.g.][]{BinneyTremaine08,HeggieHut03,Spitzer87} becomes very
complex, mostly due to the fact that the IMBH is not fixed at the centre of the
system. It becomes even more difficult when we add the emission of
GWs---another layer of complication to the Newtonian problem. As of now, we
must rely on computer simulations to address this problem.

The joint detection of a GW source with different observatories has been
already discussed in the literature but not in the mass ratio range that is
addressed in this work.  The series of works
\citep{ASF06,Amaro-SeoaneEtAl09a,AS10a,Amaro-SeoaneSantamaria10} investigated
the formation, evolution, inspiraling and merger of IMBH binaries with a mass
ratio not larger than 10 and the prospects of multiband detection with LISA and
LIGO/Virgo. The work of \citep{KocsisLevin2012} explored a joint detection by
different GW detectors in more detail than the previous references in the
context of bursting sources emitted by binaries in galactic nuclei, also with a
mass ratio not larger than 10. After the first detections of LIGO, the prospect
for the detection of similar-mass ratio stellar-mass black holes with masses
of about $30\,M_{\odot}$ with LIGO/Virgo and LISA was discussed in
\citep{Sesana2016}, and \citep{ChenAmaro-Seoane2017} clarified that this is
only possible for eccentric binaries in that mass rage.

In this paper I show that IMRIs, typically forming in globular clusters, but
without excluding larger systems such as galactic nuclei and dense nuclear
clusters, can be jointly detected with ground-based observatories and
space-borne ones.  In particular, the advanced Laser Interferometer
Gravitational-Wave Observatory (LIGO) and Virgo, and the proposed third
generation Einstein Telescope \citep{SathyaprakashEtAl2012,HildEtAl2011}, will
be able to detect IMRIs from very eccentric and hard binaries, which form via
two-body relaxation or the parabolic capture of a compact object and abrupt loss of energy.
This idea was first presented in the work of \cite{QuinlanShapiro1989},
while the energy and angular momentum changes in the case of a hyperbolic
orbit were presented previously in \cite{Hansen1972}, and see
\cite{KocsisEtAl2006,MandelEtAl2008,OlearyEtAl09,LeeEtAl2010,HongLee2015} for more recent
works.
LISA however is deaf to these kind of sources. For milder eccentricities and
semi-major axis, however, the combined detection with LISA and LIGO/Virgo or the
ET of IMRIs is a real possibility.  Due to the range
of frequencies that these sources have, a decihertz observatory such as the
DECi-hertz Interferometer Gravitational Wave Observatory
\citep{KawamuraEtAl2011}, the Superconducting Omni-directional Gravitational
Radiation Observatory \citep[SOGRO, see][]{PaikEtAl2016,HarmsPaik2015} or the
proposed geocentric Tian Qin \citep{LuoEtAl2016} would enhance the prospects of
detection.

For some systems, LISA can give advance warning to ground-based detectors weeks
before the source appears in their bandwidth and with an accuracy of seconds
(and possibly below) before the merger.

\section{Formation of Intermediate-mass ratio inspirals in globular clusters}

In this work the sources of interest are inspirals of compact objects on to an
IMBH with a mass ratio of about $\sim 10^2-10^4:1$.  The most accurate
simulations of a globular cluster are the so-called direct-summation $N-$body
algorithms. In this scheme, one directly integrates Newton's equations of
motion between all stars in a cluster at every timestep, with a regularisation
algorithm for binaries, so that any phenomenon associated with gravity naturally
arises \citep[see e.g.][and the latter for the concept of
regularisation]{Aarseth99,Aarseth03,AarsethZare74}.  Following the first
implementation of \citep{KupiEtAl06}, many modern direct-summation codes can
mimic the effects of general relativity via a post-Newtonian expansion of the
forces to be integrated \citep[see section 9 of][for a review of
stellar-dynamical relativistic integrators]{Amaro-SeoaneLRR}.

The first dynamical simulation that presented the formation and evolution of an
IMRI down to a few Schwarzschild radii from coalescence using this scheme is
the work of \cite{KonstantinidisEtAl2013}. In one of the simulations we
presented, we observed and tracked the spontaneous production of an IMRI
between an IMBH of mass $M_{\rm BH}=500\,M_{\odot}$ and a stellar-mass black
hole of mass $m_{\rm CO}=26\,M_{\odot}$. After a few Myrs, the IMRI merges and
the IMBH receives a relativistic recoil
\citep{CampanelliEtAl2006,BakerEtAl2006,GonzalezEtAl2007} and escapes the whole
cluster. It must be noted that the IMBH was in a binary for almost all of the
simulation time with another compact object, a stellar-mass black hole. The
IMBH exchanged companions a few times and was ionised for a last time very
abruptly to form the last binary. This binary started at a very small
semi-major axis, of about $a \sim 10^{-5}$ pc, and a very large eccentricity,
of $e=0.999$, which fits in the parabolic capture mechanism of
\citep{QuinlanShapiro1989}. A few years later, \cite{LeighEtAl2014} find similar
results for a close range of masses but with a different approach.  The work of
\cite{HasterEtAl2016} follows very closely the initial setup of
\cite{KonstantinidisEtAl2013} and reproduces our results with a different
integrator, which corroborates our findings. Last,
the numerical experiments of \cite{MacLeodEtAl2016} explore IMBHs in a lighter
range, of masses around $M_{\rm BH}=150\,M_{\odot}$. They however also report
that the IMBH forms a binary for about 90\% of the time. The probability
distribution of semi-major axis peaks at about $\lesssim 10^{-5}$ pc.

\section{Light and Medium-size IMRIs}
\label{sec.light}

The characteristic amplitude and the GW harmonics in the quadrupolar
radiation approximation can be calculated following the scheme of
\cite{PM63}, in which the orbital parameters change slowly due to
the emission of radiation. This is emitted at every integer multiple of
the orbital frequency, $\omega_n=n\,\sqrt{G\,M_{\rm BH}/a^3}$, with $a$
the semi-major axis. The strain amplitude in the n-th harmonic at a given
distance $D$, normalized to the typical values of this work is

\begin{align}
    h_n &= g(n,e) \frac{G^2\,M_{\rm BH} m_{\rm CO}}{D\,a\,c^4} \\
        \nonumber &\simeq  8\times 10^{-23} g(n,e)
    \left(\frac{D}{500\,\mathrm{Mpc}}\right)^{-1}
    \left(\frac{a}{10^{-5}\,\mathrm{pc}}\right)^{-1} \nonumber \\
    & \left(\frac{M_{\rm BH}}{10^3\,M_{\odot}}\right)
    \left(\frac{m_{\rm CO}}{10\,M_{\odot}}\right).
\end{align}

\noindent
In this expression $M_{\rm BH}$ is the mass of the IMBH, $m_{\rm CO}$ is the
mass of the compact object (CO), and $g(n,\,e)$ is a function of the harmonic
number $n$ and the eccentricity $e$ \citep[see][]{PM63}. We consider the RMS
amplitude averaged over the two GW polarizations and all directions. Other
alternatives to this approach, such as the works of
\cite{PPSLR01,GlampedakisEtAl2002,BarackCutler2004,GG06} give a more accurate
description of the very few last orbits, but remain substantially equivalent to
\cite{PM63} at previous stages of the evolution. This approach gives a correct
estimation of the frequency cutoff at the innermost stable circular orbit
(ISCO) frequency and is enough for the main goal of this work \citep[and see
the work of][for a discussion about the detection of binaries with mass ratios
of 0.1 with advanced ground-based detectors using aligned-spin
effective-one-body waveforms]{VeitchEtAl2015}.

\begin{figure*}
\resizebox{\hsize}{!}
          {\includegraphics[scale=1,clip]{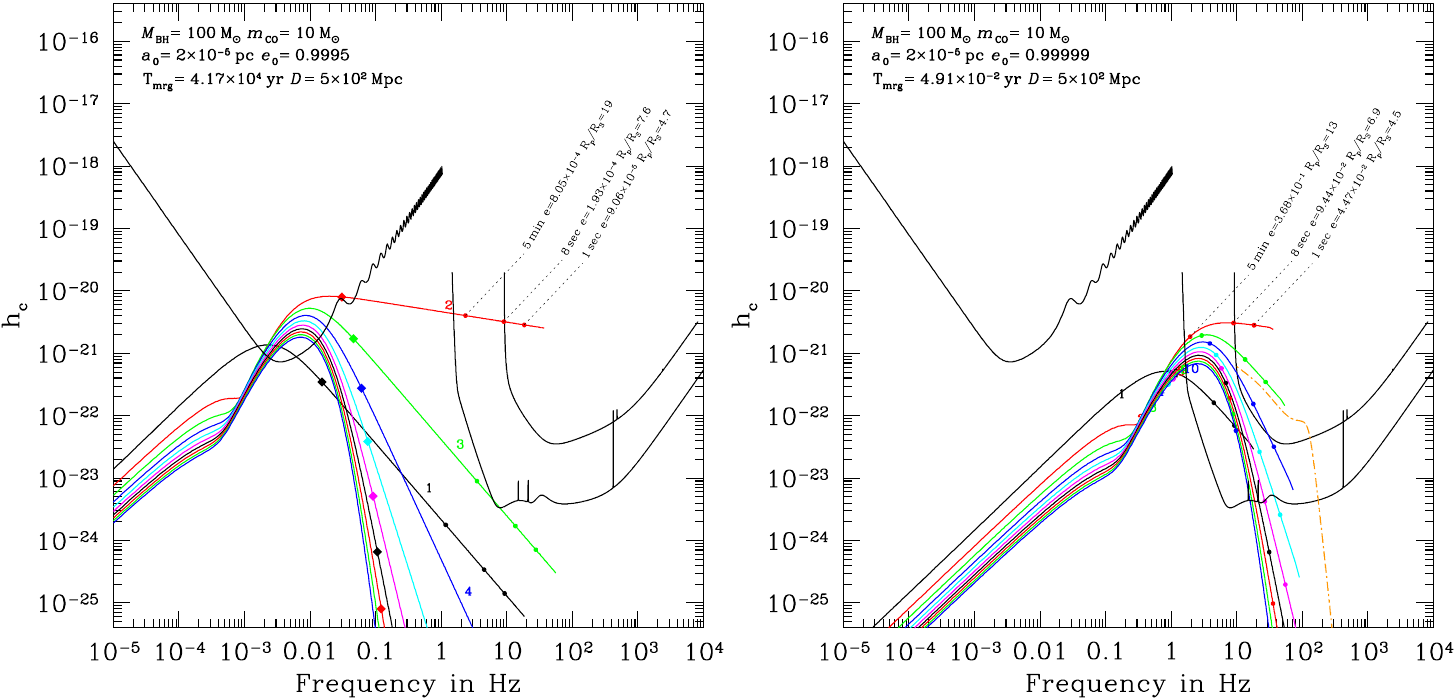}}
\caption
   {
Characteristic amplitude of the first ten harmonics emitted during the
evolution of an IMRI. The left, solid V-shaped curve corresponds to LISA's
intrinsic noise, and the two right U-shaped curves to the ET (lower $h_c$
values) and to Advanced LIGO. The mass of the IMBH is fixed to $M_{\rm
MBH}=100,\,M_{\odot}$ and the mass of the compact object is $m_{\rm
CO}=10\,M_{\odot}$. The source is assumed to be located at a distance of
$D=500\,$Mpc. Each panel corresponds to a binary with different initial values
for the semi-major axis or eccentricity. I localise and show on the second
harmonic a few instants of time in the evolution of the binary before the final
merger. The total amount of time for the binary to merge from the initial
values of the semi-major axis and eccentricity is given in each panel, $T_{\rm
mrg}$. The square symbol corresponds to one year before it. The rest of the
harmonics also display the same instants of time using the same symbol but
without a text label. I show the value of the eccentricity
in that particular moment and the pericentre value $R_{\rm p}$ in function of
the Schwarzschild radius $R_{\rm S}$. Additionally, I depict in the right panel
with a dashed, orange curve the full waveform of the system in the LIGO sensitivity curve
as approximated by the IMRPhenomD algorithm presented in \citep{HusaEtAl2016,KhanEtAl2016}.
   }
\label{fig.100}
\end{figure*}

With this approximation, I show in Fig.~(\ref{fig.100}) $h_{\rm c}$ as function
of the frequency of two different IMRIs, and a few moments in the evolution
before the final merger, which happens at a time $T_{\rm mrg}$. For the kind of
eccentricities that I am considering in this work, this time can be estimated
following \citep{Peters64} for typical values as

\begin{align}
 T_\mathrm{mrg} &\cong \frac{24\sqrt{2}}{85} \frac{(1-e_0)^{7/2} c^5}
 {G^3 M_\mathrm{BH}^2 m_{\rm CO}} a_0^4 \cong
 6.4\times 10^{5} \mathrm{yrs} \\
 & \nonumber \times
 \left(\frac{M_{\rm BH}}{10^3\,M_{\odot}}\right)^{2}
\left(\frac{m_{\rm CO}}{10\,M_{\odot}}\right)^{-1}
\left(\frac{R_\mathrm{P}^0}{200\,R_{\rm S}}\right)^{4}\nonumber \\
& \left(\frac{1-e_0}{10^{-5}}\right)^{-1/2},
\label{eq.Tmrg}
\end{align}

\noindent
where $R_\mathrm{P}^0$ and $e_0$ are the initial pericenter distance and
eccentricity, respectively.  In this Fig.~(\ref{fig.100}) the IMBH has a mass
of $M_{\rm BH}=100\,M_{\odot}$ and the mass of the compact object (CO) is set
to $10\,M_{\odot}$. I depict the LISA sensitivity curve and those of Advanced
LIGO (LIGO, henceforth) and the ET in its D configuration \citep{HildEtAl2011},
although I have shortened the characteristic amplitude to start at lower values
for clarity, since none of the sources I have considered achieves higher
values. For reference, I include as well the full waveform in the LIGO sensitivity curve as
estimated by the IMRPhenomD approach of \citep{HusaEtAl2016,KhanEtAl2016},
which has been developed to study sytems with mass ratios of up to $q=18$. This
curve is close to the peak of harmonics in amplitude for this specific case but
in general this is not true, and depends on the specifics of the binary
such as periastron argument, inclination angle, precession of the orbital plane, to
mention a few.

We can see that eccentricities corresponding to those that we can expect for a
dynamical capture as described in the introduction produce IMRIs which are
observable with LISA and both the ET and LIGO. In particular, the left panel
corresponds to an IMRI which spends half a minute in LIGO. For lighter masses
of the CO, this time becomes larger. For higher eccentricities, which can be
achieved via two-body relaxation or in the parabolic braking scenario, at these masses the IMRIs can be seen
only by ground-based detectors, with a significant amount of time and the vast
majority of the harmonics in band.  It is interesting to note that the ET has
been estimated to be able to detect up to several hundred events per year, see
\citep{Miller02,GairEtAl09}.

\begin{figure*}
\resizebox{\hsize}{!}
          {\includegraphics[scale=1,clip]{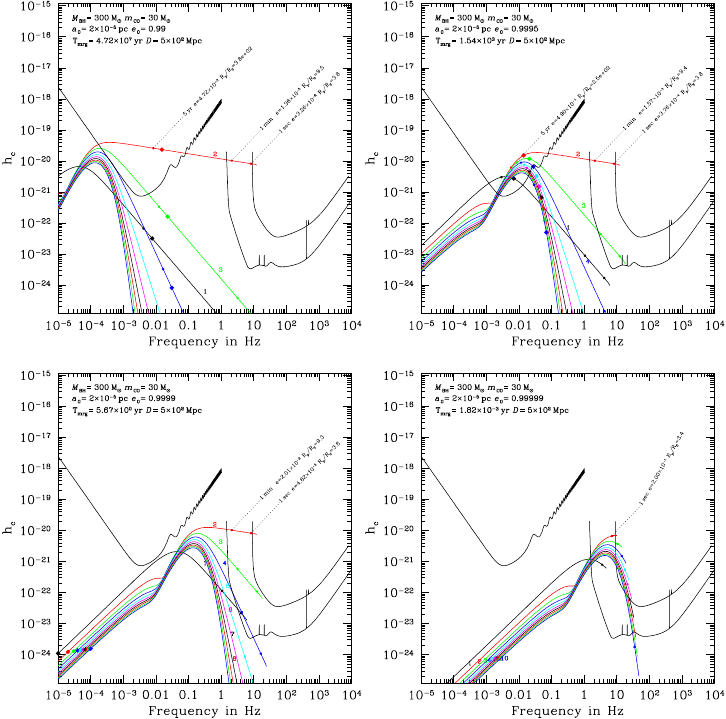}}
\caption
   {
Same as in Fig.(\ref{fig.100}) but for $M_{\rm MBH}=300,\,M_{\odot}$ and
$m_{\rm CO}=30\,M_{\odot}$ and different labels in the evolution.
Notice the displacement of the peak of frequencies, which wanders from the
LISA band to the LIGO/Virgo one (top, left panel to the bottom, right one).
   }
\label{fig.300}
\end{figure*}

In Fig.~(\ref{fig.300}) I show a more massive system, with a total mass of
$310\,M_{\odot}$. The source recedes in frequency due to the larger mass.  For
the systems considered in the upper panels, this allows IMRIs to spend more
time in LISA and accumulate more SNR, with the resulting shortened time in the
ground-based detectors which, however, is still significant. For the lower
panels, however, LISA is again deaf to these sources.

\begin{figure*}
\resizebox{\hsize}{!}
          {\includegraphics[scale=1,clip]{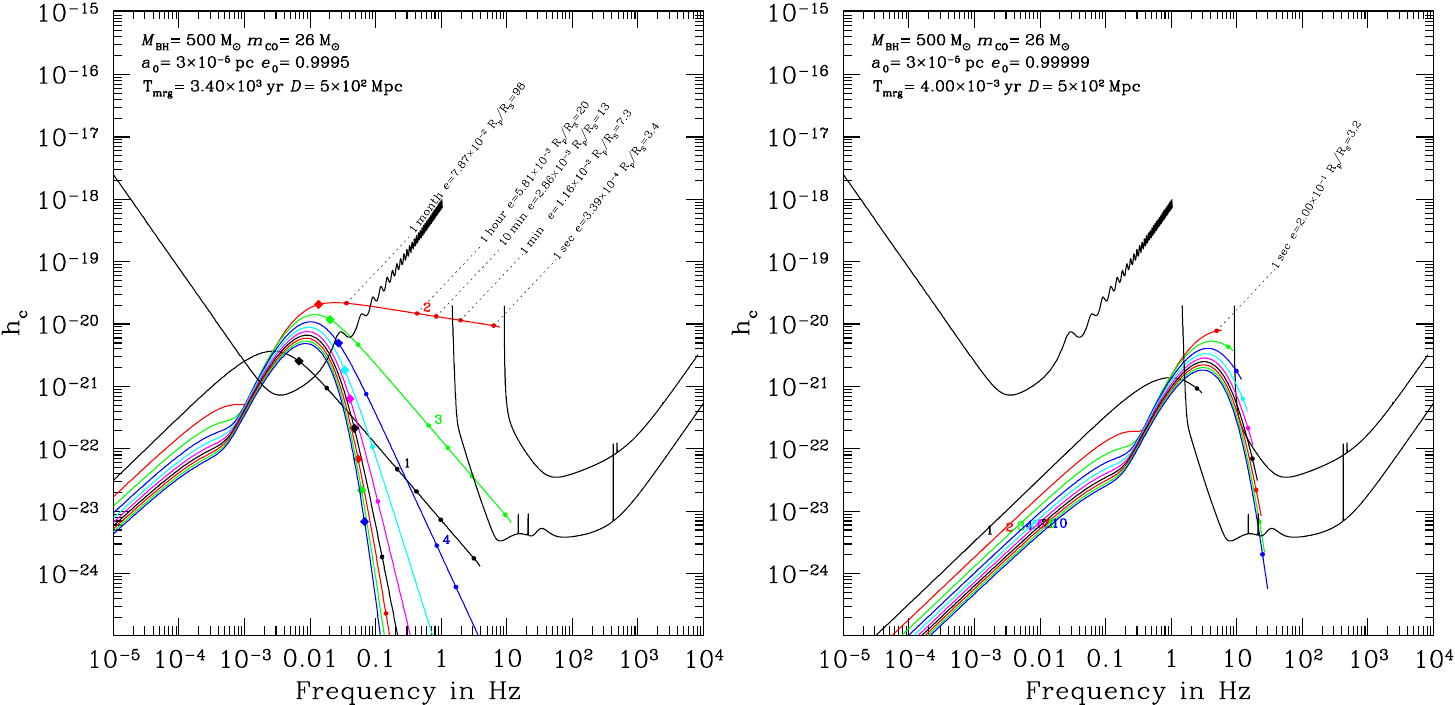}}
\caption
   {
Same as in Fig.(\ref{fig.100}) but for $M_{\rm MBH}=500,\,M_{\odot}$ and
$m_{\rm CO}=26$, which is based in the relativistic stellar-dynamical
simulation of \cite{KonstantinidisEtAl2013}. The left panel corresponds to the
kind of eccentricity in that work and the right one to a more extreme one, and
I show different labels in the evolution.
   }
\label{fig.500}
\end{figure*}

Finally, in Fig.~(\ref{fig.500}) I show a system similar to what is found in the
numerical simulations of \cite{KonstantinidisEtAl2013}. The mass of the IMBH is
set to $500\,M_{\odot}$ Higher frequencies lead the source to be observable by
only ground-based detectors.

\section{Large-mass IMRIs}

\begin{figure*}
\resizebox{\hsize}{!}
          {\includegraphics[scale=1,clip]{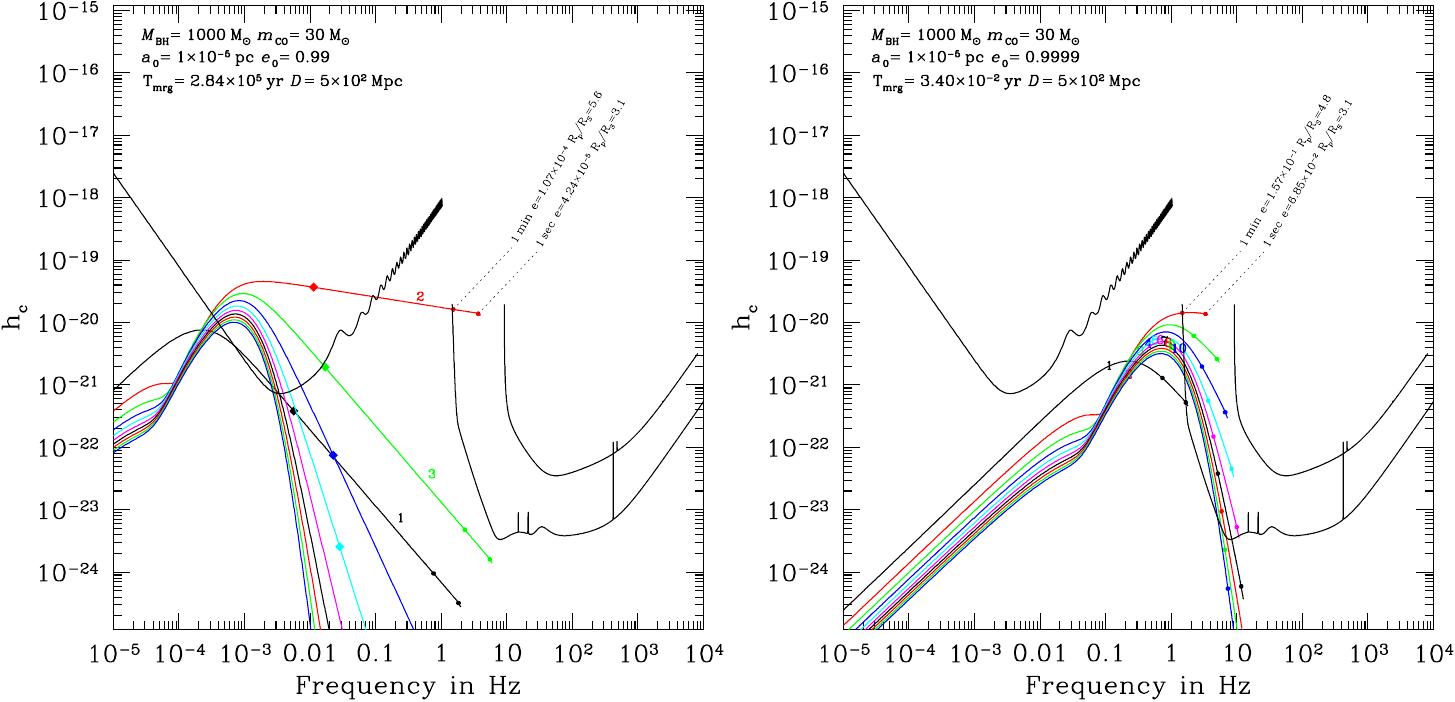}}
\caption
   {
Same as in Fig.(\ref{fig.100}) but for $M_{\rm MBH}=1000,\,M_{\odot}$ and
$m_{\rm CO}=30$ and different labels in the evolution.
   }
\label{fig.1000}
\end{figure*}

\begin{figure*}
\resizebox{\hsize}{!}
          {\includegraphics[scale=1,clip]{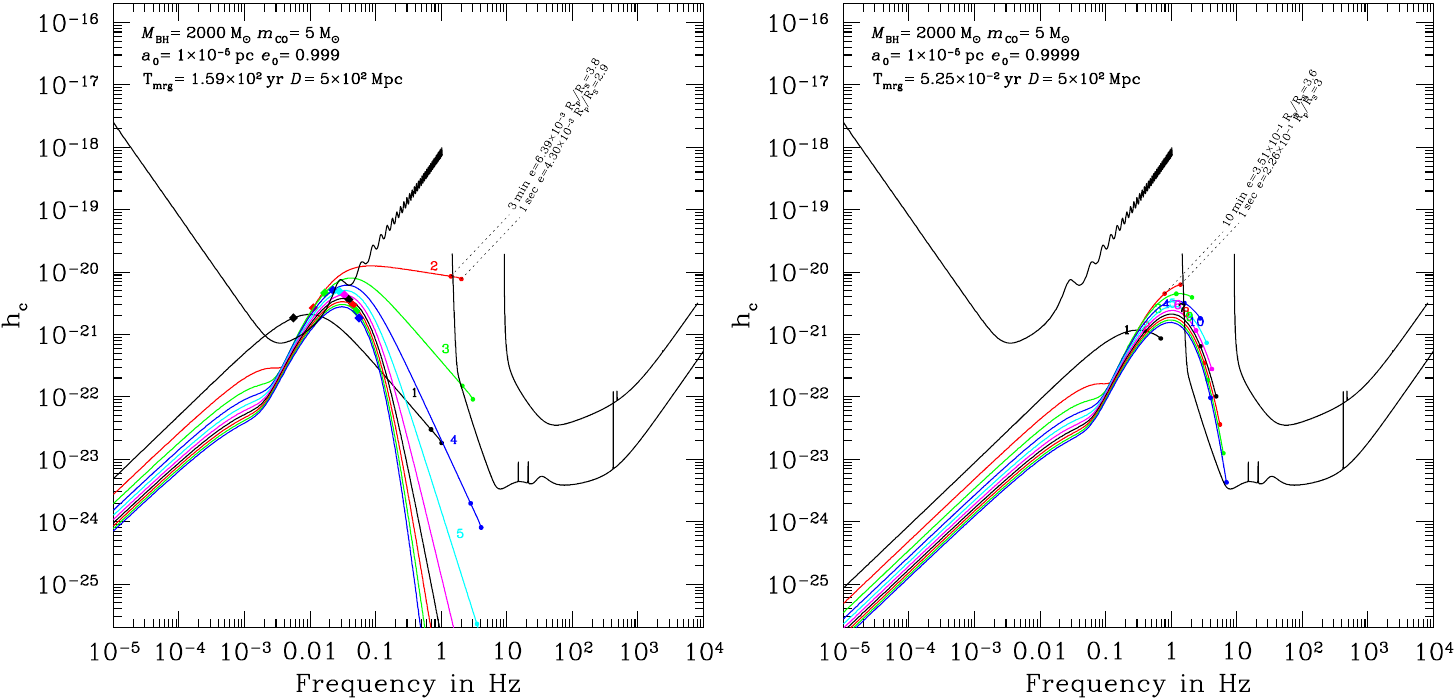}}
\caption
   {
Same as in Fig.(\ref{fig.100}) but for $M_{\rm MBH}=2000,\,M_{\odot}$ and
$m_{\rm CO}=5$ and different labels in the evolution.
   }
\label{fig.2000}
\end{figure*}

\begin{figure}
\resizebox{\hsize}{!}
          {\includegraphics[scale=1,clip]{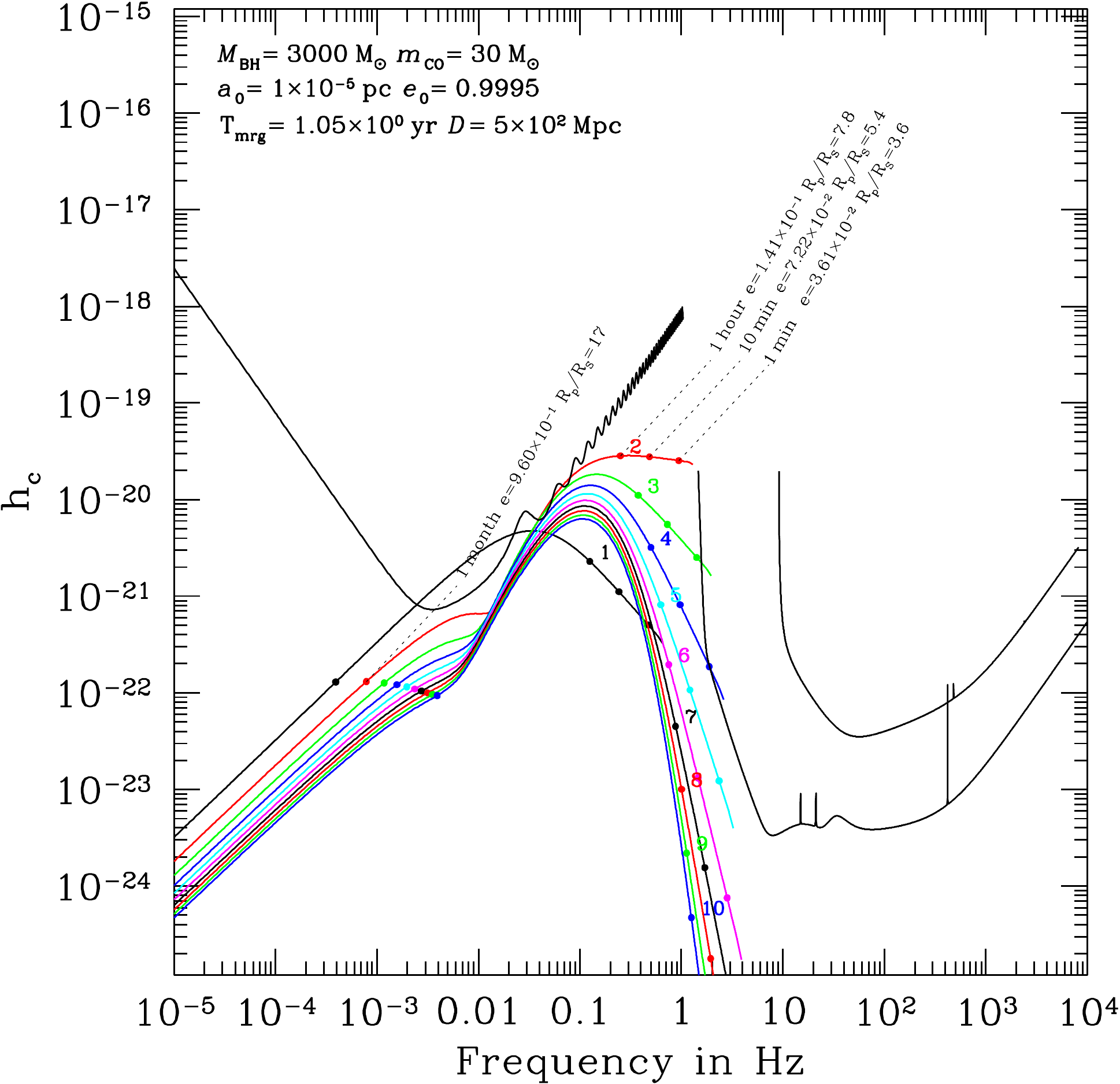}}
\caption
   {
Same as in Fig.(\ref{fig.100}) but for $M_{\rm MBH}=3000,\,M_{\odot}$ and
$m_{\rm CO}=30$ and different labels in the evolution.
   }
\label{fig.3000}
\end{figure}

In Figs.~(\ref{fig.1000}), (\ref{fig.2000}) and (\ref{fig.3000}) we can see
IMBHs with masses $M_{\rm BH}=1000\,M_{\odot}$, $2000\,M_{\odot}$ and
$3000\,M_{\odot}$, respectively.  For more moderate eccentricities, the IMRIs
in the examples can be detected with LISA and the ET, but they do not enter the
LIGO detection band. More extreme eccentricities lead to a large amount of
harmonics entering the ET band for significant amounts of time. In the case of a
$2000\,M_{\odot}$ IMBH, it can spend as much as 10 minutes in band in different
harmonics. Larger masses, i.e. $3000\,M_{\odot}$ produce short-lived sources
that however spend up to one minute in band of the ET.

\section{Environmental effects}

In the previous sections I have shown the evolution of an IMRI under the
assumption that the binary is perfectly isolated from the rest of the stellar
system. I.e. the binary evolves only due to the emission of GWs.  The reason
for this is that the problem is cleaner and easier to understand.  However, the
binary is located in a dense stellar system, typically a globular cluster.
While the role of gas is negligible, since the gas density in these systems is
very low.  Hence, so as to assess whether surrounding stars could vary or
modify the evolution \textit{after} the IMRI has formed, in this section I
investigate the impact of the stellar system in a semi-analytical approach.
The basic idea is to split the evolution of both the semi-major axis and the
eccentricity in two contributions, one driven by the dynamical interactions
with stars (subscript {\rm D}) and one due to emission of GWs (subscript {\rm
GW}), $\dot{a} = \dot{a}_{\rm GR} + \dot{a}_{\rm D}$, and $\dot{e} =
\dot{e}_{\rm GR} + \dot{e}_{\rm D}$ with dots representing the time derivative.

From \citep{Peters64},

\begin{align}
\dot{a}_{\rm GW} = &-\frac{64}{5}\frac{G^3{M}_{\rm BH}\,{m}_{\rm CO}({M}_{\rm BH}+{m}_{\rm CO})}{c^5a^3(1-e^2)^{7/2}} \\ \nonumber
                                      & \Big(1+\frac{73}{24}e^2+\frac{37}{96}e^4 \Big)\nonumber\\
\dot{e}_{\rm GW} = &-\frac{304}{15}\frac{G^3{M}_{\rm BH}\,{m}_{\rm CO}({M}_{\rm BH}+{m}_{\rm CO})}{c^5a^4(1-e^2)^{5/2}} \\ \nonumber
                                      & e\Big(1+\frac{121}{304}e^2\Big)
\end{align}

\noindent
The {GW} terms are as given in \cite{Peters64}.  Using the relationships of
\cite{Quinlan96}, we have that

\begin{equation}
\dot{a}_{\rm D}=-H\,\frac{G\rho}{\sigma}a^2.
\end{equation}

\begin{figure*}
\resizebox{\hsize}{!}
          {\includegraphics[scale=1,clip]{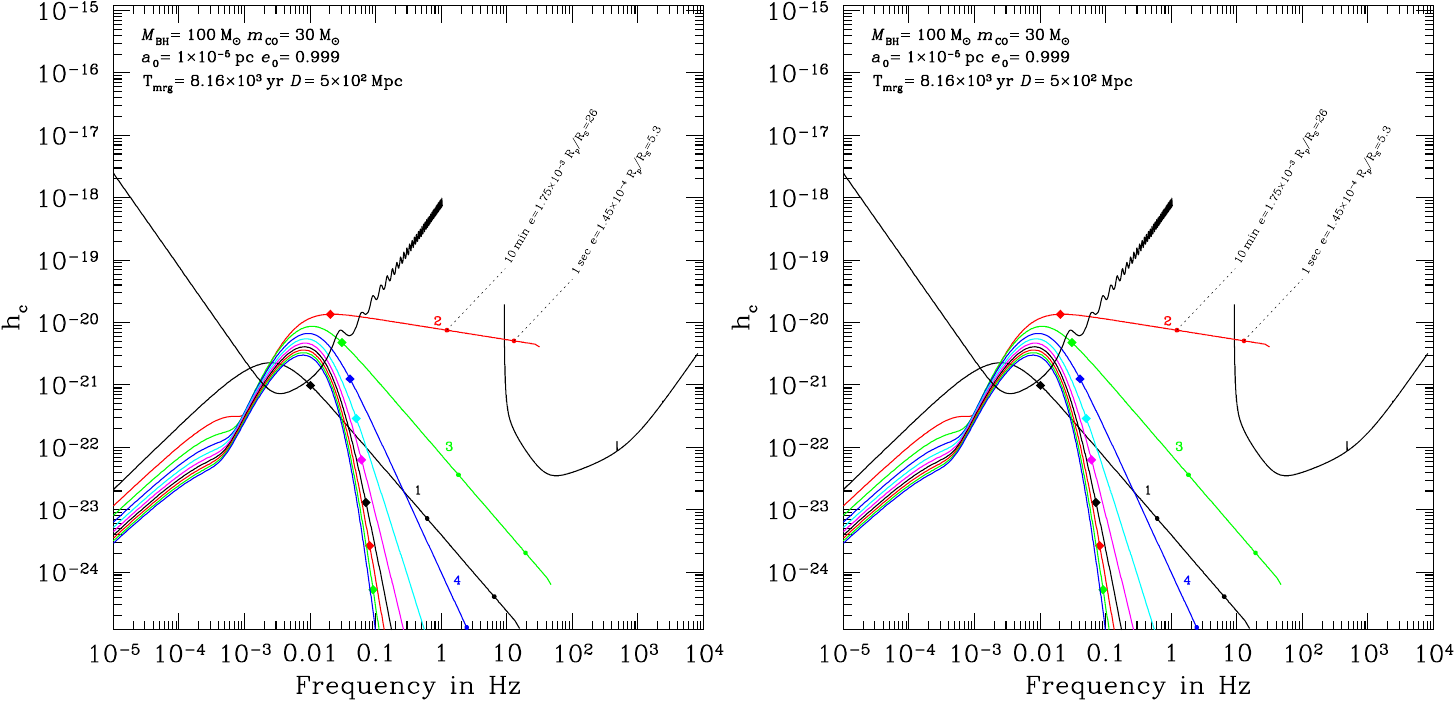}}
\caption
   {
{\em Left panel:} As in Fig.(\ref{fig.100}) but for $M_{\rm
MBH}=100,\,M_{\odot}$ and $m_{\rm CO}=30$ and different labels in the
evolution. {\em Right panel:} Same as the left one but taking into account
stellar dynamics (see text). I adopt an ambient stellar density of $2\times
10^5\,M_{\odot}\,{\rm pc}^{-3}$, $K_0=0.1$ and a one-dimensional velocity dispersion of
$\sigma=15\,{\rm km/s}$
   }
\label{fig.100-30-Vacuum-Dynamics}
\end{figure*}

\noindent
Following the usual notation, $G$ is the gravitational constant, $\rho$ is the
stellar density around the binary, $\sigma$ the corresponding velocity
dispersion of the cluster and $H$ the so-called hardening constant, as
introduced in the work of \citep{Quinlan96}.  For the kind of binaries I am
considering in this work, i.e. hard ones, we have that $\left({de}/{d\ln(1/a)}\right)_{\rm
D}=K(e)$. Since the density drops significantly during the evolution, we can
regard $\sigma$ as approximately constant and hence
$de=K(e)\,d\ln(1/a)=-{K(e)}/{a}\,da$, so that $H\simeq 16$, as in the original
work of \citep{Quinlan96} and see also \citep{SesanaEtAl04}.  Therefore,

\begin{equation}
\dot{e}_{\rm D}=\frac{H}{\sigma}\,{G\rho\,a}\,K(e),
\end{equation}

\noindent
with $K(e)\sim K_0\,e(1-e^2)$, as in the work of \citep{MM05}. As an example,
in Fig.~(\ref{fig.100-30-Vacuum-Dynamics}) I show an IMRI formed by an IMBH of
mass $M_{\rm BH}=100\,M_{\odot}$ and a CO of mass $m_{\rm CO}=30\,M_{\odot}$.
The left panel corresponds to the case in vacuum, i.e. the binary evolves only
due to the emission of GWs and the right panel takes into account stellar
dynamics.  The reason for this choice of parameters is twofold: On the one
hand, the impact of stellar dynamics on a lighter IMRI is more pronounced and,
on the other hand, $K_0$ has been estimated for more equal-mass binaries than
the other cases.  As expected, the role of stellar dynamics on to the binary at
such a hardening stage is negligible, so that the previous results hold even if
we do not take into account the surrounding stellar system around the IMRI from
the moment of formation. The previous dynamical story is however crucial for the
initial orbital parameters of the binary.

\section{Loudness of the sources}
\label{sec.SNR}

\subsection{Low-eccentricity sources: LIGO}
\label{sec.low-ecc}

As it progresses in the inspiral, a compact binary becomes observable and more
circular. The characteristic amplitude $h_{\rm c}$ of an IMRI emitting at a
given frequency $f$ is given by

\begin{equation}
h_{\rm c} = \sqrt{(2\dot E/\dot f)}/(\pi D),
\end{equation}

\noindent
with $\dot E$ the power emitted, $\dot f$ the time derivative of the frequency
and $D$ the distance to the source \citep{FinnThorne2000}.  The sky and
orientation-averaged SNR of a monochromatic source with the ansatz of ideal
signal processing is given by the equation

\begin{equation}
\left(\frac{S}{N}\right)^2 = \frac{4}{\pi D^2} \int \frac{\dot{E}}{\dot{f} \, S_h^{SA}(f)} \frac{{\rm d}f}{f^2}
\end{equation}

\noindent
as derived in \citep{FinnThorne2000}, where $D$ is the distance to the source,
$\dot{E}$ is the rate of energy lost by the source, $\dot{f}$ is the rate of
change of frequency and $S_h^{SA}(f) \approx 5 S_h(f)$ is the sky and
orientation average noise spectral density of the detector. For a source with
multiple frequency components, the total SNR$^2$ is obtained by summing the
above expression over each mode.

\begin{figure}
\resizebox{\hsize}{!}
          {\includegraphics[scale=1,clip]{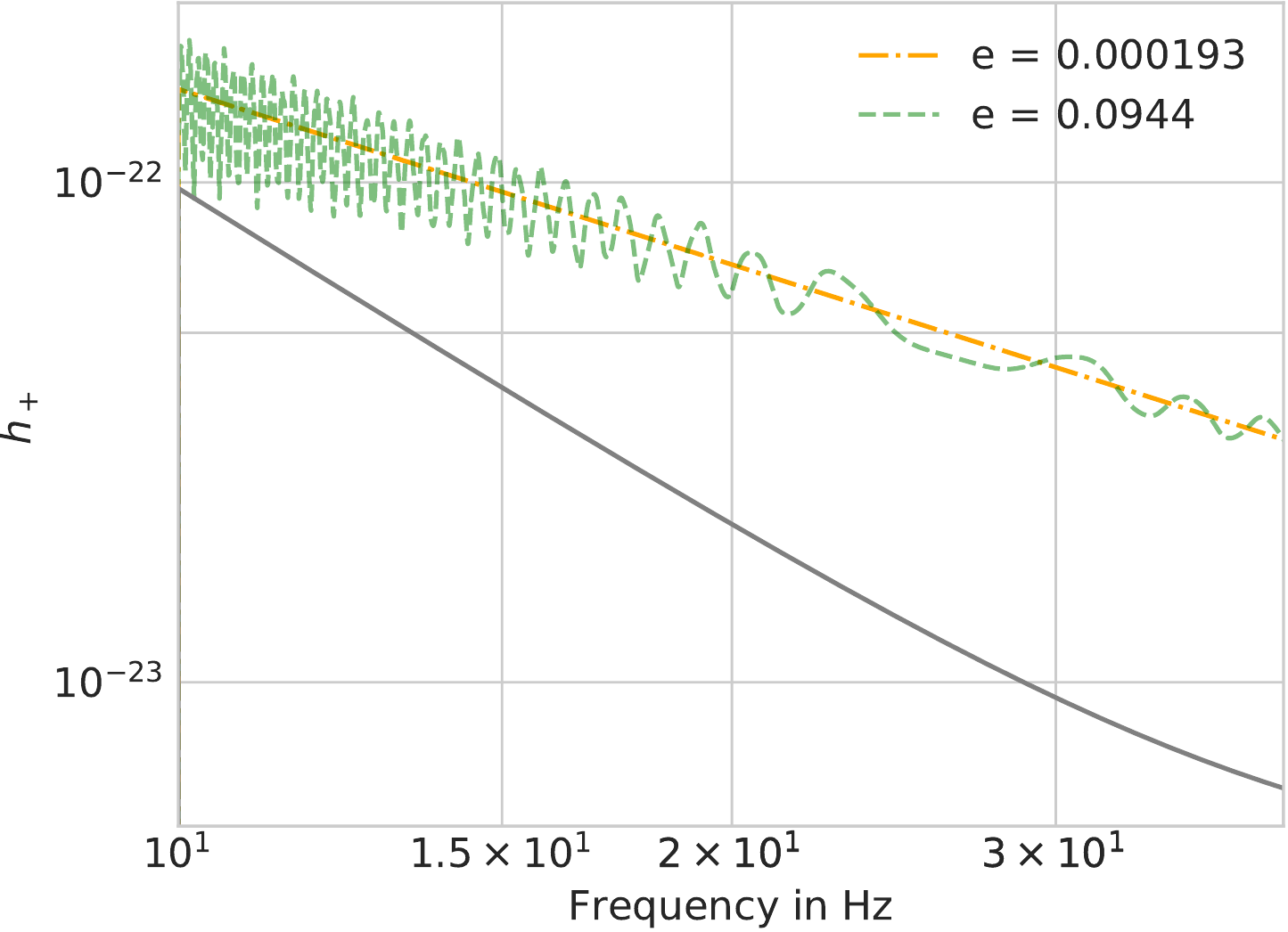}}
\caption
   {
Plus polarization $h_+$ for the two systems of Fig.~(\ref{fig.100}) from the
eccentricity of entrance in the LIGO bandwidth, as approximated by the
Fourier-transformed time domain Taylor T4 algorithm of \citep{HuertaEtAl2018},
which includes the effects of mild orbital eccentricity ($\lesssim 0.2$). The
orange, dot-dashed curve corresponds to the left panel, and the green, dashed
curve, of higher eccentricity, to the right panel of that figure, respectively.
The solid, grey curve shows the LIGO Zero Detuned High Power design
sensitivity.
   }
\label{fig.Enigma}
\end{figure}

In Fig.~(\ref{fig.Enigma}) I show the Fourier-transformed waveform of both
panels of Fig~(\ref{fig.100}), as approximated by the algorithm of
\citep{HuertaEtAl2018}. Theirs is a time-domain waveform that describes
binaries of black holes evolving on mildly eccentric orbits, not exceeding
$e\lesssim 0.2$. When the binaries enter the LIGO/Virgo band, even if they
start with initially high eccentricities, they reach values below the threshold
of the algorithm, which therefore is a good approximant to estimate
the waveform and compute the SNR.

For the IMRI examples given in Figs.~(\ref{fig.100}), assuming a distance of
$D=500\,\textrm{Mpc}$, I find a SNR in the LIGO bandwidth of 42.87 and 42.55, for
the left and right panels, respectively.  In Figs.~(\ref{fig.300}), at the
same distance, I find 17.12, 17.13 for the top-left, and top-right panels,
respectively and 17.15, 16.40 for the lower-left and lower-right ones.

\subsection{High-eccentricity sources}

When moving to lower frequencies, the eccentricity exceeds by far the limit of
the approximation of \cite{HuertaEtAl2018} that I have used to derive the SNR.
To calculate it when the IMRIs sweep the LISA bandwidth, I use
the expression (derived from Eq. 20 of \citep{PM63}, Eq. 2.1 of
\citep{FinnThorne2000} and Eq. 56 of \cite{BarackCutler2004})

\begin{equation}
\left(\frac{S}{N} \right)^2_n = \int^{f_n(\rm t_{fin})}_{f_n(\rm t_{ini})}
\left(\frac{h_{\rm c,\,n}(f_n)}{h_{\rm det}(f_n)} \right)^2
\underbrace{\frac{1}{f_n}\,d\left(\ln(f_n) \right)}_{\textrm{simply}~ df_n}.
\end{equation}

\noindent
In this Eq. $f_n(t)$ is the (redshifted) frequency of the n harmonic at time $t$
($f_n=n \times f_{\rm orbital}$), $h_{\rm c,\,n}(f_n)$ is the
characteristic amplitude of the $n$ harmonic when the frequency associated
to that component is $f_n$, and $h_{\rm det}$ is the square root of the
sensitivity curve of the detectors.

A few examples of the SNRs for the IMRI systems in the LISA band of the
previous sections (and ET in parentheses for the same source) , assuming a
distance of 500 Mpc and taking the contribution of the first 100 harmonics are:
Fig.~(\ref{fig.100}) 15 (1036), left panel, and virtually 0, 0.01 (1087) for
the right one. For Fig.~(\ref{fig.300}), the upper, left panel 50 (1994) and
the upper, right panel 24 (1995), while the lower, left panel has 2 (1991), and
the lower, right one approximately 0, 0.01 (2231). In Fig.~(\ref{fig.500}), the
left panel yields an SNR of 36 (1449), and the right one of about zero, 0.05
(1461). In Fig.~(\ref{fig.1000}), the left panel has 79 (328), and the right
one approximately zero, 0.4 (305). Fig.~(\ref{fig.2000}) has 7 (15) in the left
panel and approximately 0 in the right one, 0.1 (37). Finally,
Fig.~(\ref{fig.3000}) has 5 (1).

\begin{figure*}
\resizebox{\hsize}{!}
          {\includegraphics[scale=1,clip]{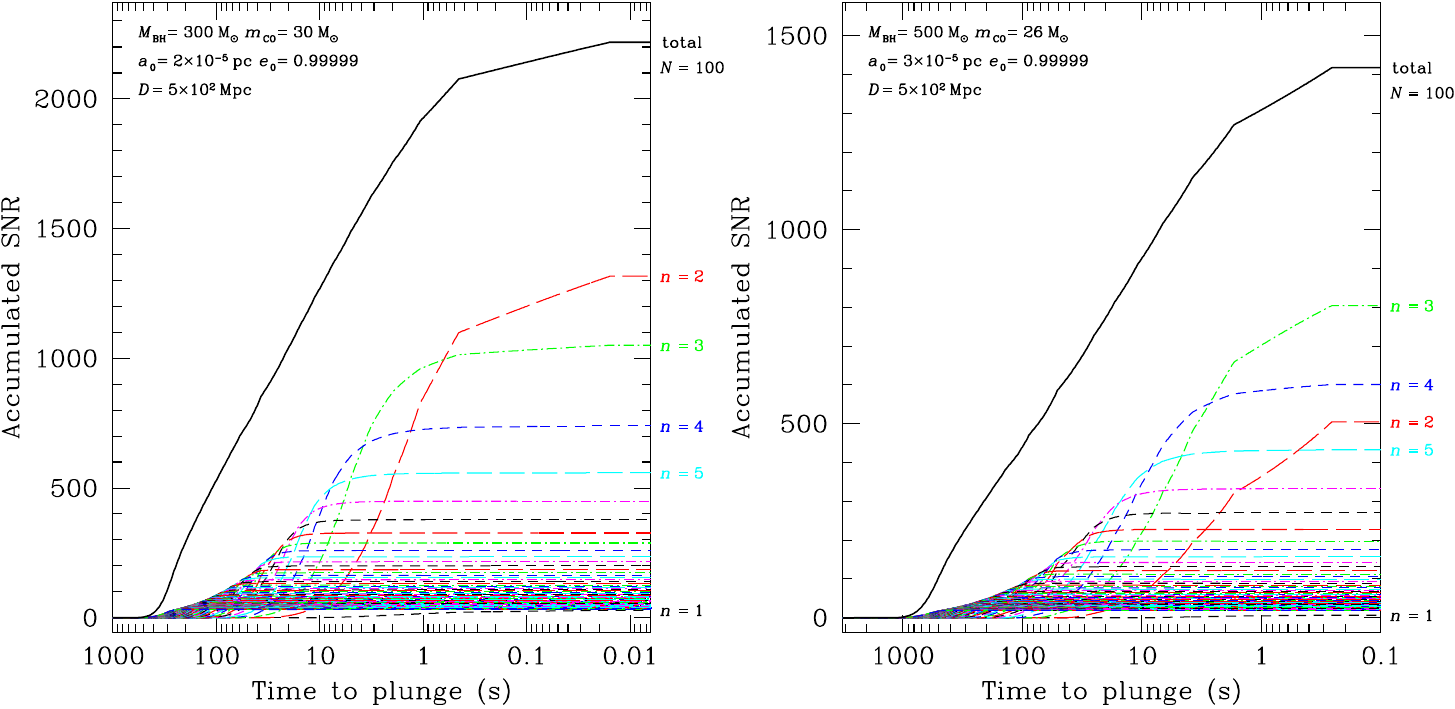}}
\caption
   {
\textit{Left panel:} Accumulated SNR in the ET as function of the time to plunge, $T_{\rm mrg}$, in seconds,
for the IMRI of Fig.~(\ref{fig.300}), bottom, right panel. I show the individual contributions
of the first 100 harmonics and the total.
\textit{Right panel:} Same for Fig.~(\ref{fig.500}), right panel.
   }
\label{fig.SNR_Fig2bottomright_and_Fig3right_ET}
\end{figure*}

\begin{figure}
\resizebox{\hsize}{!}
          {\includegraphics[scale=1,clip]{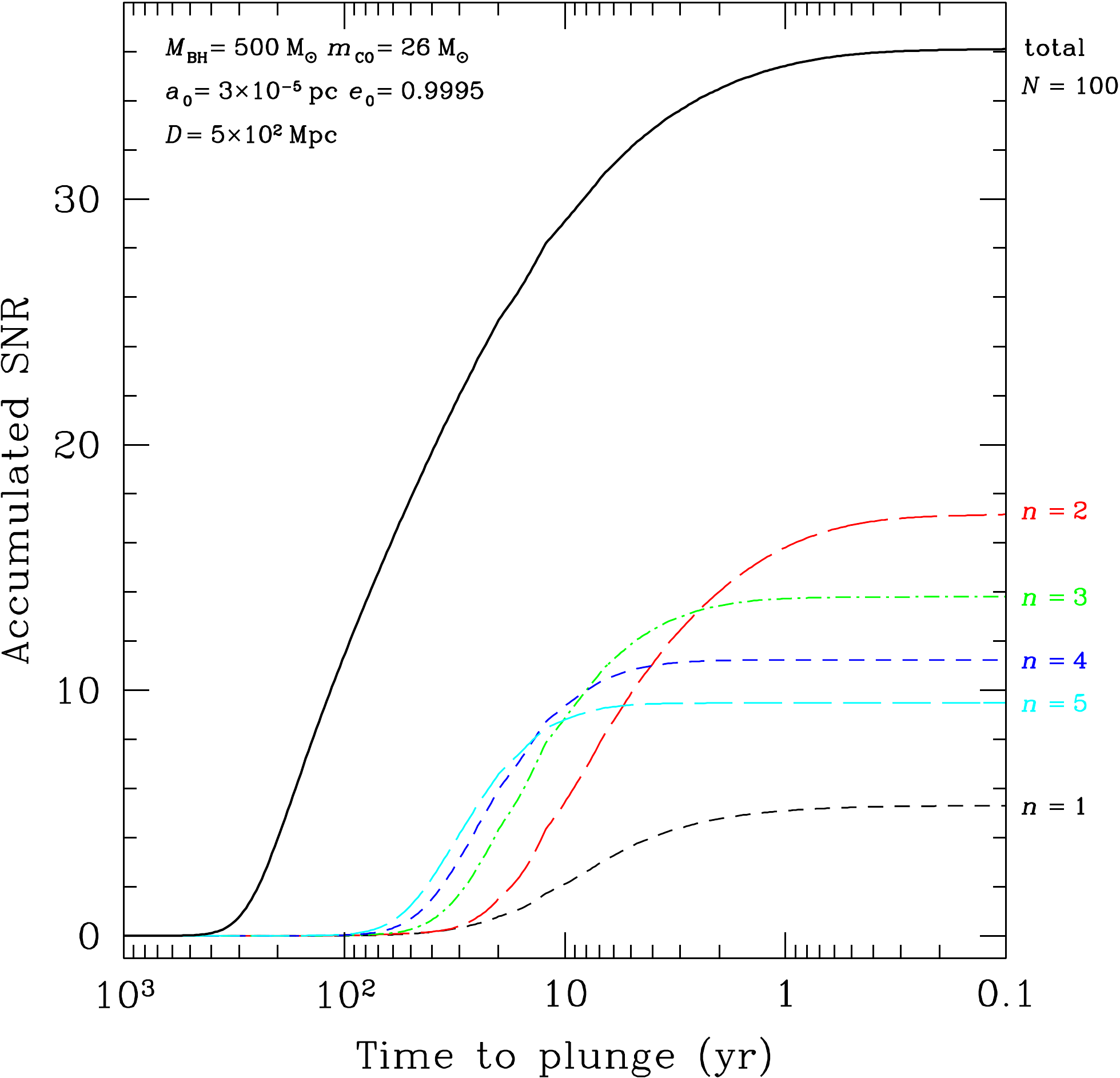}}
\caption
   {
Same as in Fig.~(\ref{fig.SNR_Fig2bottomright_and_Fig3right_ET}) but for the
same IMRI system of the left panel of Fig.(\ref{fig.500}), and in years.  I show the
individual contributions of the first 10 harmonics, but the total SNR takes
into account the contribution of the first 100, which are not displayed.
   }
\label{fig.SNR_Fig3Left_LISA}
\end{figure}

In Figs.~(\ref{fig.SNR_Fig2bottomright_and_Fig3right_ET}) and
(\ref{fig.SNR_Fig3Left_LISA}) I give three examples of the accumulated SNR as
calculated in this section. In the first figure I display in the left panel the
SNR in ET of the system of Fig.~(\ref{fig.300}), bottom, right panel and, on
the right panel, of  Fig.~(\ref{fig.500}), right panel, also for ET. In the
second one I show the accumulated SNR of the system depicted in
Fig.~(\ref{fig.500}), left panel, for LISA.

\begin{figure*}
\resizebox{\hsize}{!}
          {\includegraphics[scale=1,clip]{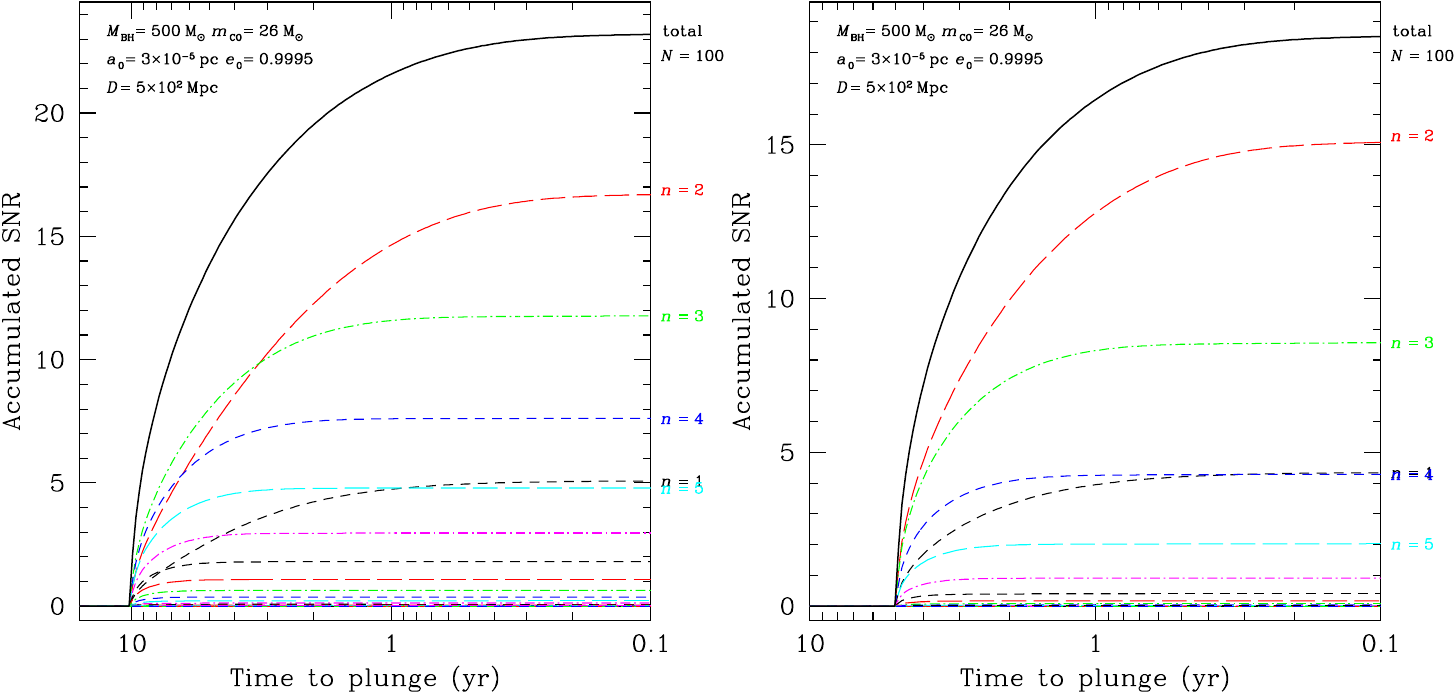}}
\caption
   {
Same as in Fig.~(\ref{fig.SNR_Fig3Left_LISA}) but taking into account only the SNR
accumulated 10 (left panel) and 5 (right panel) years before the merger. See discussion
in text.
   }
\label{fig.SNR_500_10-5-yrs-before-plunge}
\end{figure*}

However, and for the case of LISA, this is the total accumulated SNR for the
total time that the source spends on band. The observational time, the time
during which we retrieve data from the source, is in all cases shorter and,
hence, the accumulated, observed SNR is lower. As an example, for
Fig.~(\ref{fig.500}), left panel, if we integrate all of the time the source
spends on band, we obtain the aforementioned SNR of 36. However, if we
integrate the last 10 yrs before merger, the SNR goes down to 23, and to 19 for
the last 5 yrs. If we observed the source earlier in the evolution, say, e.g.
10 yrs before merger to 5 yrs before it, the SNR would be 14 and 100 yrs before
merger to 95 yrs, 3. I show an example for the accumulated SNR for this system
in Fig.~(\ref{fig.SNR_500_10-5-yrs-before-plunge}), 10 and 5 yrs before the final
plunge. This only applies to LISA, because the time spent on the ground-based
detector ET is much shorter.

So as to assess whether this approach is robust, I give now the SNR for the
systems of Sec.~(\ref{sec.low-ecc}) in the LIGO band, which have been
calculated with the waveform model introduced in that section. In
Fig.~(\ref{fig.100}), as estimated with this approach, the SNR is 41 and 40,
for the left and right panels, respectively. In Fig.~(\ref{fig.300}) I find,
from left to right, top to bottom, 12, 12, 11 and 14. These results are very
close to those of Sec.~(\ref{sec.low-ecc}). The small differences arise from
the fact that eccentricity tends to enhance the amount of energy emitted during
the inspiral as the system radiates in band for longer. It is reasonable to
take these estimates for circular orbits as a guideline for eccentric systems
of similar mass to these. If the source is eccentric, since $a=R_{\rm
per}/(1-e)$, $a$ is larger at the time the source reaches a frequency of 10 Hz.
The inspiral time depends on the value of $a$, and is larger for larger $a$.
Another way to see this is that $dE/dt$ is smaller when $e$ is larger at fixed
periapsis (or frequency in our approximation). This is because at fixed
periapsis, increasing the eccentricity puts more of the orbit further from the
MBH and hence the energy flux is on average reduced. As $dE/dt$ is smaller, it
takes longer to inspiral. This also explains why the SNR is slightly lower --
$dE/dt$ is lower at fixed periapsis and thus at fixed frequency in this
approximate model (physically, energy is being radiated out of band so we do
not detect it all).

\section{Accumulated phase shift}

Understanding how IMRIs form and what are their orbital parameters can help us
to reverse-engineer the environmental properties of the host cluster.  Although
the IMRIs considered in this work have very large initial eccentricities, when
they reach the LIGO/Virgo band the eccentricity is virtually zero.  It is
however important to measure a non-zero eccentricity, because it can be a
constraint on the formation mechanism as well as the stellar enviroment of the
IMRI. If a residual eccentricity is present, it will induce a difference in the
phase evolution of the signal as compared to a circular inspiral. Thanks to the
derivation of \citep{KrolakEtAl1995} of the phase correction due to non-zero
eccentricities, we can estimate the accumulated phase shift to lowest
post-Newtonian order and to first order in $e^2$ with

\begin{align}
\Delta \Psi_{e}(f) & = \Psi_{\rm last} - \Psi_{\rm i} \cong - \Psi_{\rm i} =\nonumber \\
                   &  \frac{7065}{187136}\,e_i^2\left(\pi\,f\,M_{\rm z} \right)^{-5/3}.
\label{eq.Psi}
\end{align}

\noindent
In the last equation $e_i$ is the eccentricity at the frequency of the dominant
harmonic at which it enters the detector bandwidth, $f$ is the frequency for
the $n=2$ harmonic, and I have introduced the quantity $M_{\rm z}:= (1+z)
G\left( M_{\rm BH} \times m_{\rm CO}\right)^{3/5} (M_{\rm BH}+m_{\rm
CO})^{-1/5}/c^3$.  Also, I make the approximation that $\Delta \Psi_{e}(f) =
\Psi_{\rm last} - \Psi_{\rm i} \simeq -\Psi_{\rm i}$, with $\Psi_{\rm last}$
and $\Psi_{\rm i}$ the final and initial phase.  This is so because of the
pronounced fall-off of $\Psi_{e}(f)$ with increasing frequency, see discussion
in section B.2 of \cite{CutlerHarms2006}.

So as to derive the accumulated phase shift in terms of $f$ and the remaining
time to merger, we now recall from \citep{Kepler1619} that the semi-major axis
of the binary is

\begin{equation}
a^3 = \frac{G\left(M_{\rm BH}+m_{\rm CO} \right)}{\left(\pi\,f\right)^2}.
\label{eq.K}
\end{equation}

The time for merger for $e \ll 1$ can be derived from \citep{Peters64} as follows,

\begin{align}
T_{\rm mrg} & \cong \frac{5}{256} \frac{c^5}{G^3M_{\rm BH} \times m_{\rm CO}
\left(M_{\rm BH}+m_{\rm CO} \right)} \nonumber \\
            & \left[\frac{G(M_{\rm BH}+m_{\rm CO})}{(\pi\,f)^2} \right]^{4/3}.
\label{eq.TmrgLowEcc}
\end{align}

Last, let us recall that

\begin{equation}
e^2\,f^{19/9} \cong \textrm{constant},
\label{eq.efconst}
\end{equation}

\noindent
which can be derived from relation 5.12 of \cite{Peters64} with $1/(1-e^2) \simeq 1$
combined with Eq.~(\ref{eq.K})\footnote{``Sed res est certissima
exactissimaque quod proporti{$\rm \bar{o}$} qua est inter bin{$\rm \bar{o}$}rum qu{$\rm \bar{o}$}rumcunque
Planet{$\rm \bar{a}$}rum tempora periodica, sit praecise sesquialtera proportionis medi{$\rm \bar{a}$}rum
distanti{$\rm \bar{a}$}rum (...)''} , i.e. $a \propto f^{-2/3}$.

Therefore, if we use Eq.~(\ref{eq.K}) in Eq.~(\ref{eq.TmrgLowEcc}), we obtain

\begin{equation}
\pi f \cong \left( \frac{5}{256} \right)^{3/8} M_{\rm z}^{-5/8} T_{\rm mrg}^{-3/8}.
\label{eq.pif}
\end{equation}

Hence, using Eqs.~(\ref{eq.Psi}, \ref{eq.efconst}, \ref{eq.pif}), we have that
the accumulated phase shift in terms of $f$, $e_i(f)$, $M_{\rm z}$ and $T_{\rm
mrg}$ is

\begin{align}
\Delta \Psi_{e}(f) & = \left(\frac{5}{256}\right)^{-17/12}\frac{7065}{187136} \nonumber \\
                   & \left(\pi f_i \right)^{19/9}e_i^2 M_{\rm z}^{25/36} T_{\rm mrg}^{17/12} \nonumber \\
                   & \cong 10 \left(\pi f_i \right)^{19/9}e_i^2 M_{\rm z}^{25/36} T_{\rm mrg}^{17/12}
\end{align}

The accumulated phase shift is detectable if $\gtrsim \pi$. With this
approximation, I find the following phase shifts in radians, for the IMRI
systems presented in the previous sections, imposing a minimum threshold SNR of
5 (the numbers correspond to the panels of the figures from the top to the
bottom, left to right):

(i) For LISA, and taking into account only the last five years before merger,
Fig.~(\ref{fig.100}) has a negligible phase shift.  Fig.~(\ref{fig.300}) 180,
$3.4\times 10^6$, while the other two panels have a a negligible phase shift.
Fig.~(\ref{fig.500}) $1.5\times 10^6$ and the right panel is negligible.
Fig.~(\ref{fig.1000}) 8200 and the right panel is negligible.
Fig.~(\ref{fig.2000}) $9.7\times 10^5$ and the right panel is negligible.
Least, Fig.~(\ref{fig.3000}) has also a negligible phase shift.

(ii) For the ET, Fig.~(\ref{fig.100}) $\sim 5.1\times 10^{-3}$, 19000 for the
left and right panels. Fig.~(\ref{fig.300}) $\sim 2.6\times 10^{-7}$, $\sim
3.4\times 10^{-3}$, 0.66 and 4600. Fig.~(\ref{fig.500}) $1.3\times 10^-3$ and
3900.  Fig.~(\ref{fig.1000}) $3.5\times 10^{-6}$ and 450. Fig.~(\ref{fig.2000})
$1.3\times 10^{-2}$ and 2600. Fig.~(\ref{fig.3000}) has a negligible phase
shift.

(iii) For LIGO, Fig.~(\ref{fig.100}) $4\times 10^{-6}$ and 1.2.
Fig.~(\ref{fig.300}) $1.1 \times 10^{-10}$, $1.4 \times 10^{-6}$, $2.3 \times
10^{-4}$ and 10. The rest of the cases have negligible phase shifts.

\section{Conclusions}

Intermediate-mass ratio inspirals are typically formed in dense stellar systems
such as galactic nuclei and globular clusters, with typically very large
eccentricities (from $e=0.999$) and small semi-major axis (below $a \sim
10^{-5}$pc), as found in a number of stellar-dynamics simulations of globular
clusters
\citep{KonstantinidisEtAl2013,LeighEtAl2014,HongLee2015,MacLeodEtAl2016,HasterEtAl2016}.
Besides classical two-body relaxation, an interesting way of explaining the
formation of these sources is the parabolic capture mechanism described by
\citep{QuinlanShapiro1989,KocsisEtAl2006}.

In this work I show that IMRIs in clusters are detectable not only by
space-borne observatories such as LISA. Depending on the properties of the
IMRI, it can be detected in conjunction with LIGO/Virgo or the ET, so that
ground-based and space-borne observatories should be envisaged as one
instrument if they are simulataneously operative.

I have considered IMBHs with masses ranging between $M_{\rm BH}=100\,M_{\odot}$
up to $3000\,M_{\odot}$ and COs with different masses. I have separated them in
light and medium-size IMRIs, for IMBHs with masses up to $500\,M_{\odot}$
(which is a particular case based on the findings of
\citep{KonstantinidisEtAl2013}) and large-mass IMRIs, for masses between
$1000\,M_{\odot}$ and $3000\,M_{\odot}$.

I find that light and medium-size IMRIs can be observed by LISA and
ground-based detectors for eccentricities starting at $0.99$ and up to
$0.9995$. In the range of frequencies of LIGO/Virgo they spend a maximum of
about one minute on band. Higher eccentricity sources, however, can only be
detected by ground-based detectors (see \citep{ChenAmaro-Seoane2017} for a
discussion on the role of eccentricity for low mass ratio binaries).  This is
due to the fact that, as the eccentricity increases, the pericenter distance
decreases, so that the characteristic frequency of the GWs emitted at the
pericenter increases (see \citep{Wen2003}, Eq. 37 for a derivation of the peak
frequency in the same approximation used in this work). In some cases, the full
cascade of harmonics falls entirely in the bandwidth of the ground-based
detectors.

The peak of large-mass IMRIs recedes in frequency as compared to light and
medium-size ones, so that the cascade of harmonics is shifted towards the LISA
domain. However, for eccentricities below $0.9995$, IMRIs with IMBHs covering
the full range of masses considered in this work ($100\,M_{\odot}$ up to
$3000\,M_{\odot}$) should be detectable with LISA with modest to large SNRs,
from a few to tens, depending on the eccentricity and duration of the
observation. For ground-based detectors, I compute the SNR for LIGO using the
waveforms from a Fourier-transformation of the time domain Taylor T4 algorithm
of \citep{HuertaEtAl2018} (limited to eccentricities $\lesssim 0.2$) and derive
large enough SNRs, always of a few tens.

Lower-frequency sources require larger eccentricities, and we cannot use these
waveforms. For these detectors I use an approximate scheme to
calculate the SNR, and I have compared it with the previous results for LIGO
and I find that the approach is robust. The values for ET can reach as much as
$\sim 2000$, and are of typically a few hundred and of tens for masses up to
$2000\,M_{\odot}$. LISA has SNRs of a few tens to then significantly drop when
the IMRI system has the peak of harmonics closer to the ground-based regime.

By combining ground-based and space-borne observations we can impose
better constraints on the system's parameters. On the one hand, LISA can
observe the inspiral and hence provide us with measurements of parameters such
as the chirp mass. On the other hand, ground-base detectors detect the merger
and ringdown, and therefore measure other parameters such as the final mass and
spin. Thanks to this joint detection, one can split various degeneracies and
get better measurements of the parameters, as compared to individual
detections\footnote{Christopher Berry, personal communication.}.

I have estimated with a semi-analytical approach the possible influence of the
environment {\em after} their formation and I find no impact, which will make
it easier to detect and interpret these sources.

By looking at the accumulated phase shift, one could investigate the origin of
light IMRIs thanks to a residual eccentricity. I find that LISA binaries
accumulate typically hundred of thousands and up to millions of radians, while
ET binaries can accumulate up to 19000 radians, and typically of a few
thousands. While IMRI binaries in LIGO live much shorter time, there is a case
which does accumulate enough radians.

LISA can warn ground-based detectors with at least one year in advance and
seconds of precision, so that this observatory and LIGO/Virgo and the ET should
be thought of as a single detector, if they are operating at the same time.
Until LISA is launched, the perspective of detecting IMRIs from the ground
opens new possibilities.

\section*{Acknowledgments}

I acknowledge support from the Ram{\'o}n y Cajal Programme of the Ministry of
Economy, Industry and Competitiveness of Spain, as well as the COST Action
GWverse CA16104. I thank Marc Freitag for his help in the implementation of the
SNR equations in the plotting subroutines, and for extended discussions about
the phase shift.  I am indebted with Leor Barack, Chen Xian, Bernard Schutz,
Thomas Dent, and Matthew Benacquista for general comments, with Frank Ohme for
his help with the waveforms, and with Jon Gair for discussions about SNR. This
work started during a visit to La Sapienza university in May 2018.  I thank
Roberto Capuzzo Dolcetta, Raffaella Schneider, Piero Rapagnani, Luigi Stella,
Valeria Ferrari, Paolo Pani and Leonardo Gualtieri for their extraordinary
hospitality.  In particular I thank the students who took part in my course,
because the many discussions and homework preparation led me to think about
this problem.

\end{document}